\documentclass{ws-ijmpc}
\usepackage{subfigure}
\begin{document}
\markboth{H. -I. Lee, S. -L. Wang and S. -P. Li}
{Climate Effect on Wildfire Burned Area in Alberta}

\catchline{}{}{}{}{}

\title{Climate Effect on Wildfire Burned Area in Alberta (1961 - 2010)}

\author{Hung-I Lee}

\address{Institute of Physics, Academia Sinica, 128 Sec. 2, Academia Rd., Nankang\\
Taipei , 11529,  Taiwan (R.O.C)\\
michael.billy@hotmail.com}

\author{Shih-Luen Wang}

\address{Institute of Physics, Academia Sinica, 128 Sec. 2, Academia Rd., Nankang\\
Taipei , 11529,  Taiwan (R.O.C)\\
r96222073@ntu.edu.tw}

\author{Sai-Ping Li}

\address{Institute of Physics, Academia Sinica, 128 Sec. 2, Academia Rd., Nankang\\
Taipei , 11529,  Taiwan (R.O.C)\\
spli@phys.sinica.edu.tw}

\maketitle

\begin{history}
\end{history}

\begin{abstract}
The spread and burned areas of wildfires in Alberta, Canada during a 50 year period, from 1961 through 2010 are studied here.  Meteorological factors that control the spread and burn area have been discussed for a long time.  In this paper, we analyze the temperature rise that could drastically enhance the spread and average burned area of wildfires.  A simple lattice model that mimics meteorological factors is also introduced to simulate the temperature effect on the spread and burned areas of wildfires.  The numerical results demonstrate the temperature effects on wildfires when compared to the empirical data.

\end{abstract}

\ccode{PACS Nos.: 02.50.-r, 89.75.Da, 89.40.-a}
\section{Introduction}
Wildfire is one of the main disturbances in an ecosystem and has been of much concern in many countries over the years.  Historically, it has been the cause of numerous and possibly irreversible damages with deep ecological as well as socio and economic impacts.  In some extreme cases, the whole ecosystem might be wiped out and also resulted in large numbers of human casualties.  It is therefore understood that the need for designing and developing effective ways of dealing with wildfires is constantly increasing as such phenomena appear ever more often.  In fact, the area burned annually by wildfire has always been a factor that influences policy decisions and future land use planning in countries such as the United States and Canada.  There are indeed many factors that can affect the spreading of wildfire and the area that will be burned once the fire started.  Among them, climate \cite{1} , forest type \cite{2}, wind \cite{3}, landscape \cite{4} and also human activities are some of the major factors that can influence the occurrence and spreading of wildfires.  Among these factors, climate seems to be a major factor that governs the rate of spread and burned area of wildfires.  For example, there are significantly more cases of wildfires reported during summer time when the temperature in a region is considerably higher.  Early wildfire history evidence from diverse climate regions and vegetation also suggests that ignitions and spreads of wildfires have strong relations to climate changes. \cite{5} \cite{6} \cite{7}  It is therefore of interest to investigate if the recent global warming issue could also have sizable effect on wildfire--whether higher temperatures will affect the size of the average burned area and the rate of spread of wildfires in a region. \cite{8} 

\begin{figure}[h]
\noindent{ \hspace*{-29pc} \psfig{file=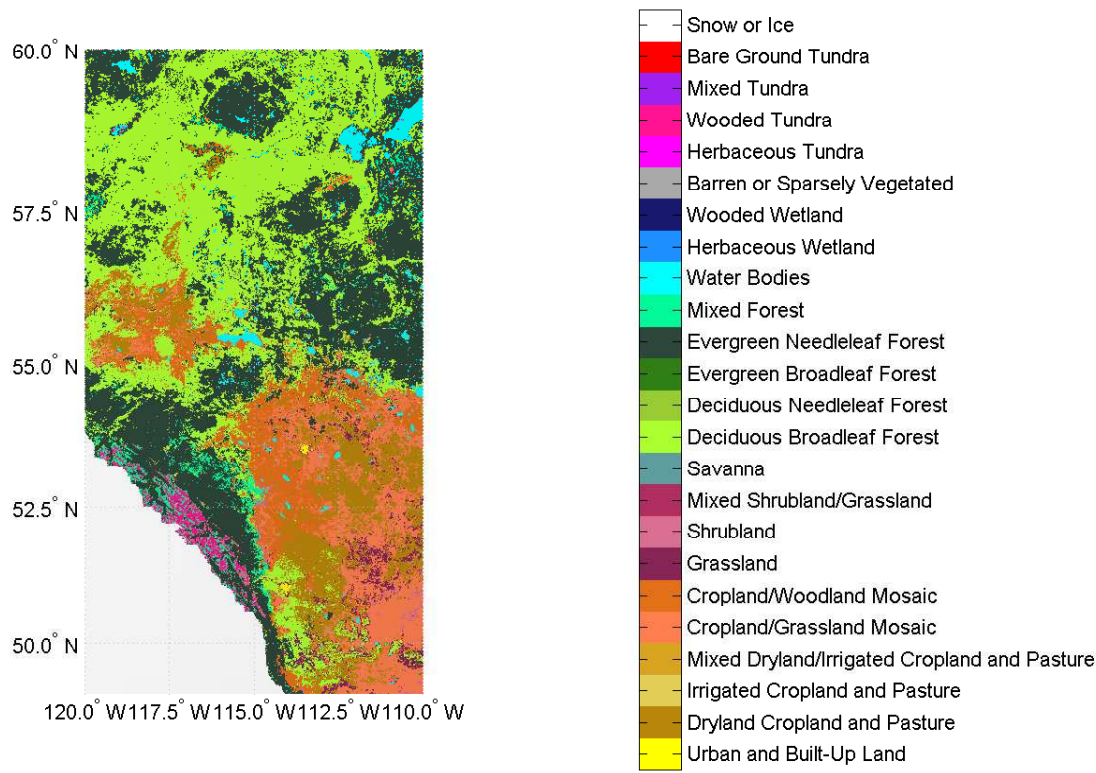,width=28cm}}
\caption{The natural ecosystems in Alberta, Canada- The land cover characterization data is derived from Advanced Very High Resolution Radiometer for an one year spanning. (from April, 1992 to March, 1993) \label{f1}}
\end{figure}

Since the recent recorded data that are available are more abundant and precise than before, one can now perform a much more accurate analysis that can help policy makers in their decisions.  There are now more wildfire data available on the web, which are accessible to the public.  The wildfire data of many regions and countries have been studied before and specific journals dedicated to the study of wildfires also exist \cite{9}. In this paper, we will analyze the historical data of the wildfires in Alberta, Canada from 1961 through 2010 \cite{10}.  We would want to make an attempt to investigate the possible temperature effect on the spreading of wildfires using data for this 50 year period.  This will hopefully shed some light on the effect of global warming on the spreading of wildfires \cite{11} and their consequences on an ecosystem.  Alberta is the fourth largest province of Canada and has a dry continental climate with warm summers and cold winters.  It is open to cold arctic weather systems from the north, which often gives extremely cold conditions in winter.  The southern and east central parts of Alberta are mainly covered by short, nutritious grass.  The trees in the parkland region of the province and on the north side of the North Saskatchewan River evergreen forests are typically aspen, poplar, and willow, prevail for hundreds of thousands of square kilometers.  This makes it an ideal region to be studied for the effects of climate changes on wildfires.  Fig.~\ref{f1} shows the natural ecosystems in Alberta derived from the database of the Advanced Very High Resolution Radiometer for a one year span. (from April, 1992 to March, 1993) \cite{12}. 

This paper is organized as follows.  In Section 2, we present an empirical study of the historical data of wildfires in Alberta for the 50 year period, from 1961 till 2010.  We then introduce a simple statistical model to analyze the empirical result obtained in Section 3.  We vary the parameter in the model to mimic the temperature effect on the dynamics of the spreading of wildfires and perform numerical simulation to compare with empirical result.  Section 4 is the conclusion of the present study.
\section{Historical Data}
The historical data of wildfires reported in Alberta are available on the web \cite{10} and we will carry out a study of these data in this section.  In the database of the website, the recorded data of a wildfire listed the date when it started, the date when it ended, the location, the total burned area and other information that are of interest for each of the reported wildfires.  For more information, the interested reader is referred to the website.  There were a total of 47,852 cases of wildfire reported during the 50 year period, from 1961 through 2010 in Alberta, Canada as listed in the database.  We notice that during this 50 year period, several changes in the recording of wildfire data were made.  For the data recorded before 1983, the burned area would be listed as 0 if it is less than 1 hectare, and for each wildfire case, the database only recorded its burned area as an integer in units of hectare.  Wildfires with smaller burned area were recorded more precisely after 1996, down to 0.1 hectare.  These changes do not affect the results and conclusions in our present study.  

\begin{figure}[h]
\centerline{\psfig{file=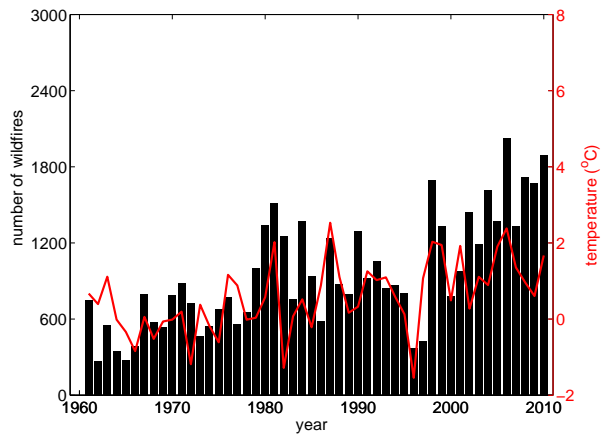,width=8cm}}
\caption{Number of wildfires reported in Alberta annually between 1961 and 2010.  The average temperature (in red) is also shown in the figure.) \label{f2}}
\end{figure}

Fig.~\ref{f2} below summarizes the number of wildfires reported in each year between 1961 and 2010.  The historical data suggested that there has been an increase in the number of wildfires reported annually starting in the late 1970's.  This increase is more significant in the first decade of the millennium.  The average temperature in the same period is also shown in the same figure in red for reference.   

\begin{figure}[h]
\centerline{\psfig{file=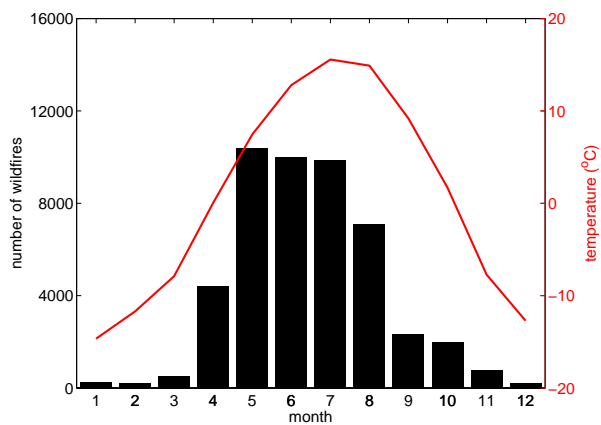,width=8cm}}
\caption{Number of wildfires vs. month and average temperature vs. month (in red). ) \label{f3}}
\end{figure}

In Fig.~\ref{f3}, we show the total number of wildfires reported in each month of the year during this 50 year period.  One can see that the months that have the most cases of wildfires are from May through August, during which the average temperatures are higher than the rest of the year.  In the same figure, we also plot the average temperature for each month during this 50 year period.  By averaging over the 50 year period, we believe that this should minimize the fluctuations in other factors such as precipitation, etc and allow us to have a better understanding of the effect of warmer climates on the spread of wildfires.  The result suggests that among possible factors, warmer climate and thus higher temperature might play an important role in the occurrence of wildfires.  

\begin{figure}[h]
\centerline{\psfig{file=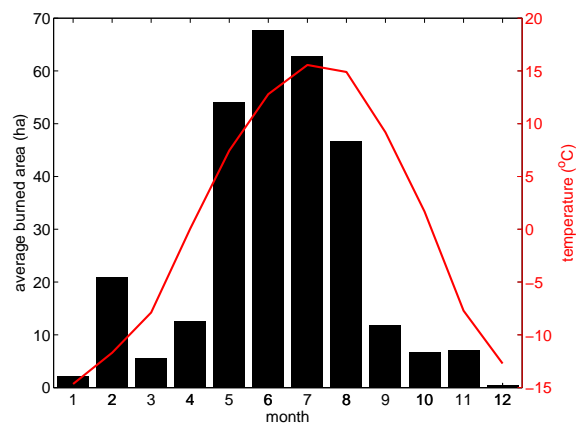,width=8cm}}
\caption{Average burned area of wildfires in each month of the year during the 28-year period from 1983 through 2010.  The average temperature (in red) is also shown in the figure.  ) \label{f4}}
\end{figure}

Fig.~\ref{f4} illustrates the average burned area of wildfires in each month of the year over the period from 1983 through 2010 for a total of 28 years.  The reason why we are unable to use data for all the years is because complete wildfire burning times were recorded only after 1983.  We also exclude a total of 12 wildfire events with burned areas larger than 50,000 (ha) because they burned for an extended period of several months and there is insufficient data to discriminate how much burned area the wildfire should contribute to each month of the burning period.  For most wildfires, the burning periods are usually short enough to be identified to what months the burning areas should contribute to.  It is easy to see that Fig.~\ref{f4} exhibits a behavior similar to that of Fig.~\ref{f3}.  The result shows that the average burned area of wildfires is much larger in these months, when the average temperature is considerably higher. 

\begin{figure}[h]
\centerline{\psfig{file=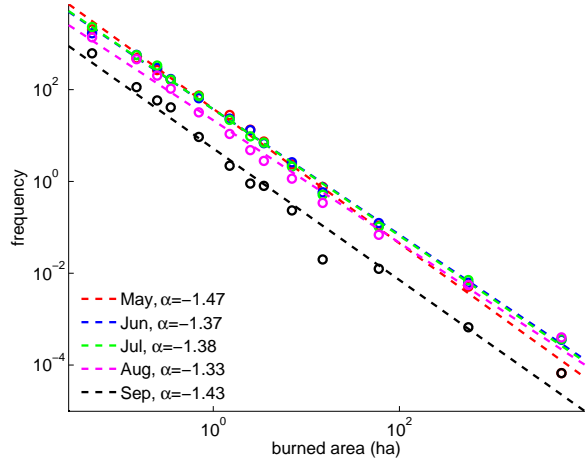,width=8cm}}
\caption{Log-log plot of the of the burned area of the reported wildfires during 1961-2010.  ) \label{f5}}
\end{figure}

Fig.~\ref{f5} is a log-log plot of the probability density function (pdf) of the burned area of the reported wildfires during this period.  We show the result for each month from May through September.  The interesting point here is that they all show heavy tail (approximate power law) behavior.  This kind of heavy tail behavior was indeed observed in the past and has been a subject of discussion \cite{13} which we will not discuss further here.    We should remark here that the exponents are slightly different for different months.  A smaller value in the magnitude of the exponent $\alpha$ would indicate more cases of wildfires with large burned areas.

\begin{figure}[h]
         \centering	
	\subfigure[\label{f6:a}]{
		\centering
		\includegraphics[width=8cm]{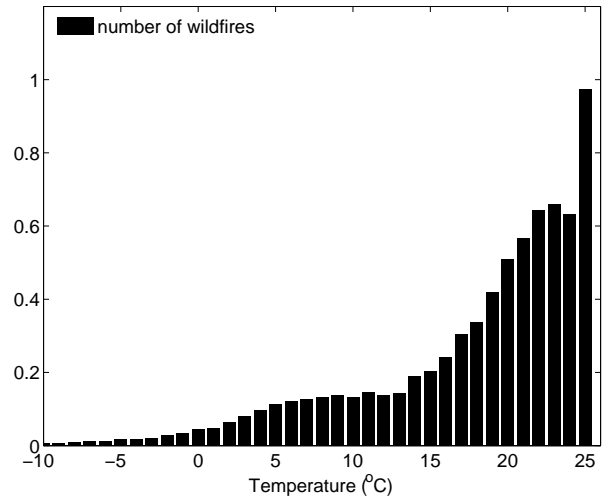}}	
	\subfigure[\label{f6:b}]{
		\centering
		\includegraphics[width=8cm]{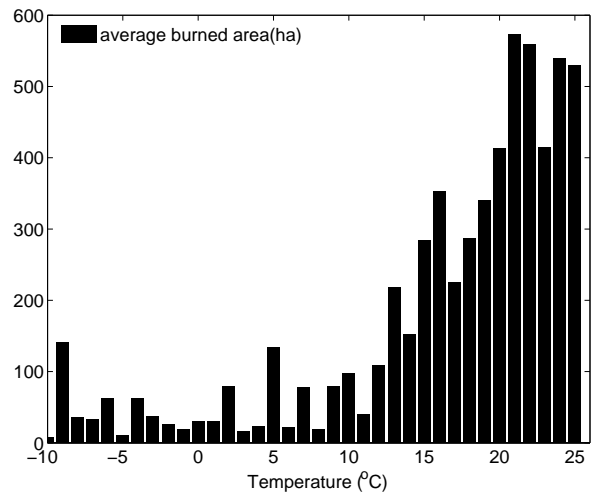}}	
	\caption{(a) Plot of the number of wildfires vs. temperature from 1961 through 2010. (b) Plot of the average burned area of wildfires vs. temperature during the period of 28 years from 1983 to 2010.  Temperature is the day when wildfires are ignited. \label{f6}}
\end{figure}

As we have mentioned earlier, it would indeed be more interesting to see how temperature can possibly affect the spread and size of the average burned area of wildfires.  Fig.~\ref{f6:a} is a plot of the number of wildfires as a function of temperature for the 50 year period and Fig.~\ref{f6:b} is a plot the average burned area of wildfires vs. temperature.  

In order to obtain the results of Fig.~\ref{f6}, we use the daily temperature as recorded in the database of the National Center for Environmental Prediction (NCEP) \cite{14}.  For our purpose, we would only need to know the qualitatively behavior of the temperature dependence.  We will therefore use the NCEP/NCAR reanalysis data \cite{14} which offer daily surface air temperature.  This temperature dataset has been tested to be reliable for diagnosing climate conditions in Canada \cite{15}.  For Fig.~\ref{f6:a}  and Fig.~\ref{f6:b} , the temperature corresponds to the temperature nearest to the location of occurrence on the day when the wildfire was first reported.  We believe that this will more likely reflect the effect of temperature on the occurrence of wildfires.  The result in Fig.~\ref{f6} suggests that both the number of wildfires and the average burned area of wildfires grow nonlinearly with respect to temperature rise.  For example, one can see that the average burned area roughly doubles as the temperature rises from $15^\circ$C to $20^\circ$C.  Likewise, the number of wildfires also roughly doubles from $15^\circ$C to $20^\circ$C.  

\begin{figure}[h]
\centerline{\psfig{file=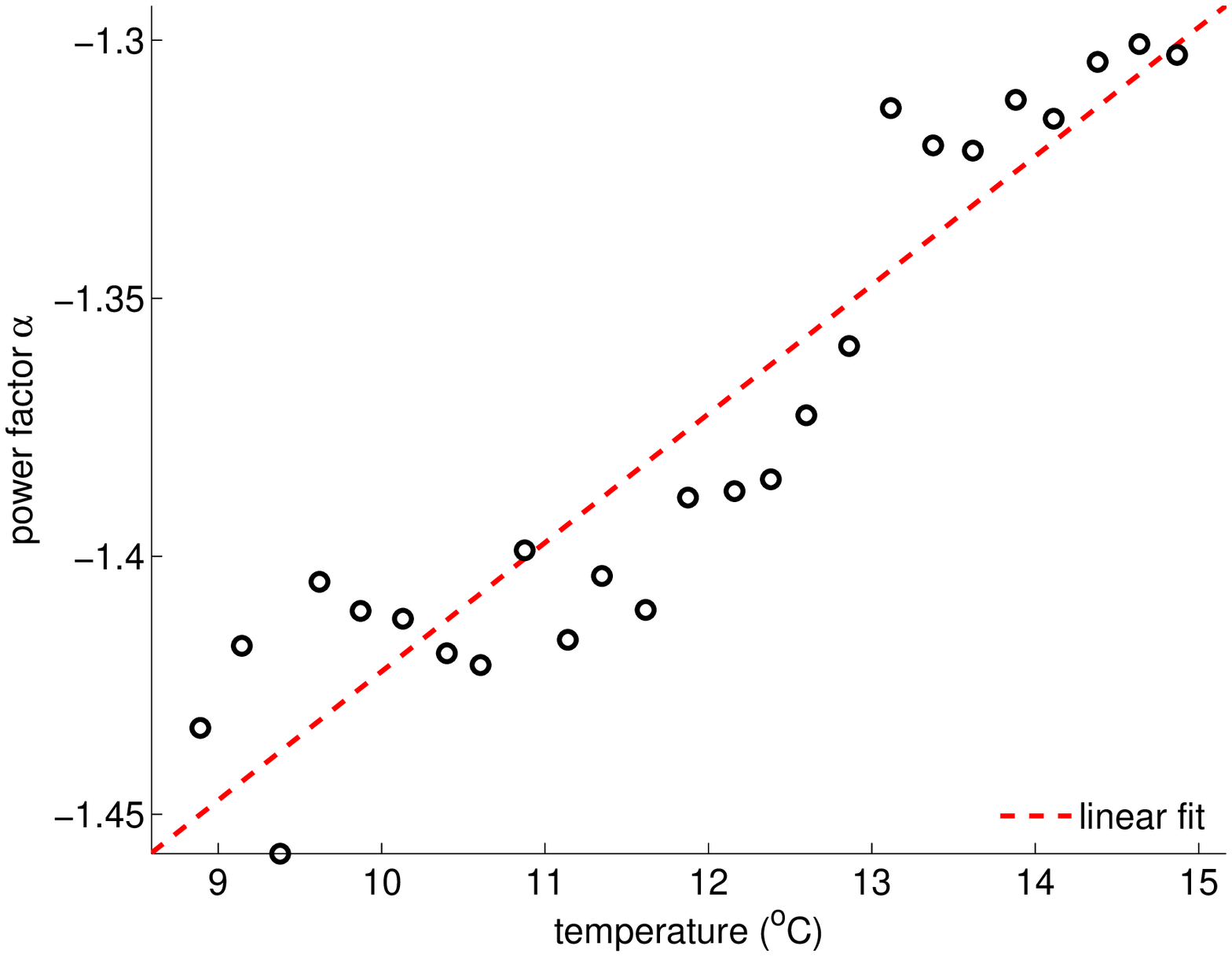,width=8cm}}
\caption{The power law exponent in Fig.~\ref{f5} as a function of temperature.  \label{f7}}
\end{figure}

In a similar fashion, one can also obtain another interesting plot, by relating the power law exponents of the pdf of burned area of wildfires as obtained in Fig.~\ref{f4} as a function of temperature.  This is illustrated in Fig.~\ref{f7}.  The result as shown in Fig.~\ref{f7} indicates that the power law exponent of the pdf of the average burned area of wildfires decreases as temperature increases.  This is consistent with the result obtained in Fig.~\ref{f5}.  It further suggests that as the temperature rises, there will be more cases of wildfires with large burned areas at higher temperatures.  Power law behaviors for wildfires have also been observed in many countries, like the United State \cite{16} \cite{17}, Australia \cite{16}, Spain \cite{18} and Italy \cite{19}. 

\section{Model Study}

To try to model wildfire dynamics by taking into account all possible factors such as temperature, precipitation, wind speed, vegetation, landscape, etc is a very complex task to achieve.  A model for predicting a wildfire spread would inevitably have to take into account these external environmental factors in a certain way.  There are already many attempts to model wildfire spreading, ranging from physical models to empirical models. \cite{20} \cite{21}  \cite{22} Each has some success but a complete picture of how to effectively describe the dynamics of wildfires is still not available for the moment.  For our purpose, we analyze the empirical data using a statistical approach.  We set an easier task by introducing a simple model to investigate the wildfire dynamics in order to understand how the effect of temperatures could possibly affect the burned area of wildfires.  This is based on a cellular automata approach by introducing a 2-dimensional lattice to mimic the forests that the wildfires would spread.  There exist works using cellular automata or lattice model approach to study wildfire in the literature, see e.g., \cite{23} \cite{24}.  To make it more specific, we here only consider a two-dimensional square lattice.  Other types of lattices such as triangular lattices, etc can be treated in a similar manner.  On each lattice site, there is only one unit of plant, or we simply say one tree/site.  There are indeed 3 states that each site can take, namely, unburned ($U$), ignited ($I$) and burned ($B$).  If the site is in state $U$, it will remain in state $U$ if it has no neighboring site in state $I$.  In our simulation, we only consider nearest neighbors.  Therefore, each lattice site will only be affected by 4 nearest neighbors in our simulation.  On the other hand, if the site has a neighboring site in state $I$, it will then have a certain probability $p$ that will be ignited and if it is ignited, it will be in state $I$ in the next time step.  Furthermore, for any cell in a state $I$, it will become burned in the next time step, i.e. in a state $B$.  Once the cell reaches state $B$, it will remain to be in state $B$ throughout the rest of our simulation and a burned site cannot ignite any of its neighboring sites once it reaches such a state.  The rules of the spreading of wildfire at each time step t can now be summarized as follows, for a lattice site $i$, its state $S_{i}(t+1)$ is given by

(1) $S_{i}(t) = U \longrightarrow S_{i}(t+1) = U$    
 (no neighboring site in state $I$ at time step $t$)

(2) $S_{i}(t) = U \longrightarrow S_{i}(t+1) = I$     (a probability $p$ from each of its neighboring site

~~~~~~~~~~~~~~~~~~~~~~~~~~~~~~~~~~~~~~~~~~ in state $I$ at time $t$)

~~~~~~~~~~~~~~~~~~$ \longrightarrow S_{i}(t+1) = U$    (a probability $1-p$ from each of its neighboring 

~~~~~~~~~~~~~~~~~~~~~~~~~~~~~~~~~~~~~~~~~~~site in state $I$ at time $t$)

(3) $S_{i}(t) = I \longrightarrow S_{i}(t+1) = B$    (the site is in state $I$ at time $t$)

\begin{figure}[h]
\centerline{\psfig{file=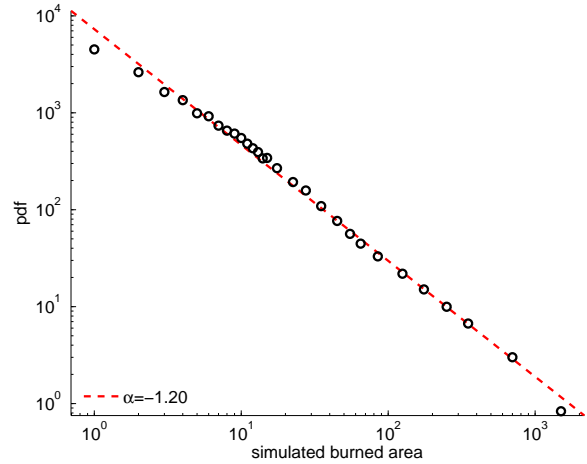,width=8cm}}
\caption{Log-log plot of pdf vs. simulated burnt area, 0.25 $< p <$ 0.55.  ) \label{f8}}
\end{figure}

From the above rules, one can easily see that for $a$ site that is in state $U$, and having $k$ 
neighboring sites in state $I$ at time $t$, the probability that it will be ignited at time $t+1$ is equal to $1 - (1-p)^k$.  In reality, the parameter $p$ will depend on many factors such as the climate, vegetation, landscape, etc of the region under study.  However, to try to relate $p$ to wildfire spreading requires one to study case by case, which is an impossible task to achieve, given that there is not enough information provided.  We therefore take the assumption that in a certain local region, and with the same vegetation and landscape, the parameter $p$ will be influenced more by the local conditions, such as the local temperature.  It is also conceivable that during the spread of the wildfire, there will be fluctuations in many factors that can affect the rate of spread and the probability of ignition.  We therefore assume that in the simulation, the wildfire spreads at a certain temperature $T$.  During the simulation, the parameter $p$ can take any value within a range, i.e. $p_{min} < p < p_{max}$.  At time step $t$, one would randomly choose a value of $p$ within this range.  One then check if the site will become ignited by its neighboring sites by performing a Monte Carlo simulation using this value for $p$.  By setting $p$ to be within a certain range at a certain temperature, we are statistically mimicking the fact that during the spread of the wildfire, there will be local fluctuations in factor such as wind, landscape, and also the temperature itself.  Thus, we shall obtain numerical results that can be used to compare with empirical data.  It is also easy to understand that the smaller values of the range of $p$ ($p_{min}$ and $p_{max}$) correspond to smaller probability for the site to be ignited, thus correspond to lower temperature.  We would here want to see if this simple model can give qualitatively and semi-quantitatively the behaviors of wildfires in a statistical way that are consistent with the empirical data.  Fig.~\ref{f8} is a log-log plot of the simulated result of the pdf (probability density function) of the burned area of wildfires on our two-dimensional square lattice, when $p$ is between 0.25 and 0.55. In the simulation, we put this on a 256 x 256 square lattice and we performed a total of 30,000 trials. In each trial, we first start with a value of p, within the preset $p_{min}$ (0.25) and $p_{max}$ (0.55). After a certain number of time steps (we
use $\sim$30 in our simulation), we pick another value of p between $p_{min}$ (0.25) and $p_{max}$ (0.55) and continue the simulation and so on until the simulated wildfire stops. One can see that the pdf of the simulated burned area of wildfires indeed has a heavy tail on the right side which can be approximated by a power law with exponent $\sim$1.2.  

\begin{figure}[h]	
         \centering	
	\subfigure[\label{f9:a}]{
		\centering
		\includegraphics[width=8.5cm]{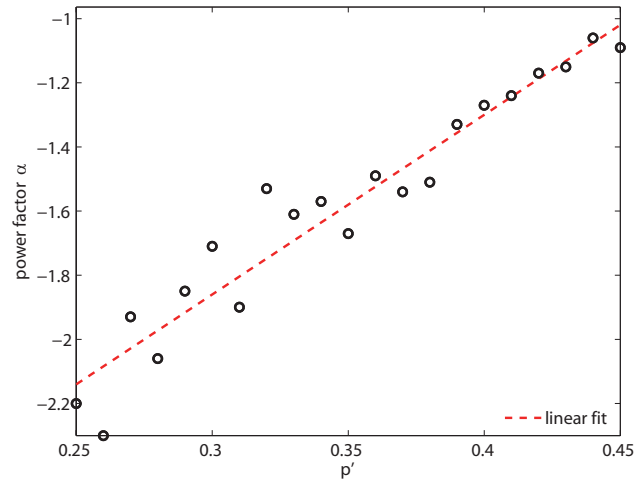}}	
	\subfigure[\label{f9:b}]{
		\centering
		\includegraphics[width=8cm]{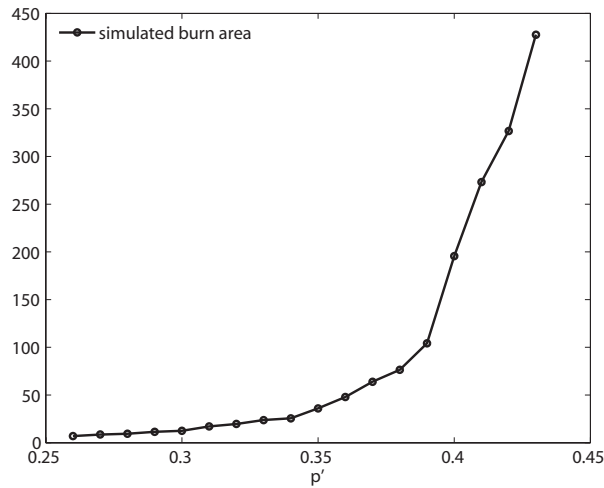}}		\caption{(a) The exponent vs. p' for the simulated burned area.  The broken red line is a linear fit with slope $\sim$5. 5 (b) Average simulated burned area vs. $p'$ \label{f9}}
\end{figure}

As we change the values of $p_{min}$ and $p_{max}$, we will obtain a different value of power law exponent for the heavy tail of the simulated burned area for different values of $p_{min}$ and $p_{max}$.  For the time being, let us choose the range of $p$ to be fixed, i.e., $p_{max} - p_{min}$ to be a constant.   To facilitate our discussion, let us define $p'$ to be the mid-point of the range of p, i.e. $p' = (p_{min} + p_{max})/2$.  We then perform simulations for different $p'$.  We again perform the simulation on a 256 $\times$ 256 square lattice and we run for a total of 30,000 trials for each $p'$. In each trial, we first start with a value of p, between the preset $p_{min}$ and $p_{max}$. After a certain number of time steps (we use $\sim$30 in our simulation), we pick another value of $p$ between $p_{min}$ and $p_{max}$ and continue the simulation and so on until the simulated wildfire stops. We set the range to be $p' \pm$ 0.15 for the simulation performed here. The result is illustrated in Fig.~\ref{f9:a}, which is a plot of the exponent vs. $p'$ for the simulated burnt area. It can be approximated by a linear fit with slope $\sim$5.  In Fig.~\ref{f9:b}, we give the result of the average simulated burned area with respect to different values of $p'$.

One can compare the simulated result of Fig.~\ref{f9:b}, to that of the empirical data
shown in Fig.~\ref{f6:b}. Both plots show similar behavior, indicating that the burned
areas increase nonlinearly as the temperature (or $p'$) increases. As a remark, we have also performed simulation using ranges between $\pm$ 0.1 and $\pm$ 0.15 for $p'$.  The result indicates that the qualitative behavior is the same but the slope changes slightly for different values within the range.

\begin{figure}[h]
\centerline{\psfig{file=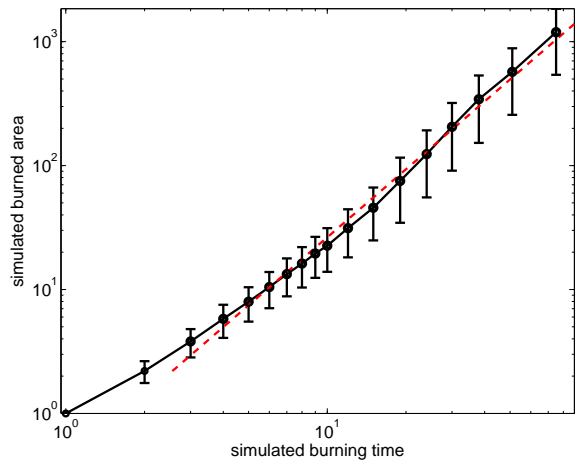,width=8cm}}
\caption{A log-log plot of the average burned area vs. average burning time needed for the range 0.25 $< p <$ 0.55.  The broken red line is a linear fit with slope $\sim$1.8.  ) \label{f10}}
\end{figure}

Using this cellular automata model, one can perform numerical study of other interesting phenomena in wildfire activities.  For example, one can perform a numerical study of the average burned area vs. average burning time needed to burn the area with respect to different $p'$.  Fig.~\ref{f10} is a log-log plot of the average burned area vs. average burning time needed for the range 0.25 $< p <$ 0.55.  Notice that the time needed grows as roughly as a power law with exponent $\sim$1.8.  For a different set of $p_{min}$ and $p_{max}$, it will correspond to an exponent of different value.  If this is what would happen in the real world, it should definitely be an issue for policymakers to be concerned with.  The numerical result shown in Fig.~\ref{f10} should in fact be used to compare with empirical data when more data are accumulated in the future.

\section{Conclusion}
In this paper, we have studied the historical data of the wildfires reported in Alberta, Canada for a 50 year period from 1961 through 2010.  We note that the wildfire activities as reported have increased since the late 1970's.  To further understand this increase of wildfire activities, we here investigate the effect of temperature change on the spread and burned areas of wildfires during this period.  To look for such an effect, we try to minimize the effect from other factors by averaging over the 50 year period of the historical data, and obtain the average burned area of wildfires in Alberta with respect to temperature change.  The result from the empirical data indicates that the average burned area grows nonlinearly as temperature rises.  It also suggests that the number of wildfires that have large burned areas also increases nonlinearly as temperature increases.  We introduce a simple lattice model to mimic the spreading dynamics of wildfires.  The parameter introduced in this model can be related to temperature.  We then perform Monte Carlo simulations on a two dimensional square lattice that mimics the spreading of wildfires.  The numerical result obtained is shown to be consistent with the result from the empirical data, suggesting that this model can be used as a first step to study the dynamics of the spread of wildfires.


\end{document}